\documentstyle[preprint,aps,epsfig]{revtex}

\newcommand{\bea}{\begin{eqnarray}}
\newcommand{\eea}{\end{eqnarray}}
\newcommand{\be}{\begin{equation}}
\newcommand{\ee}{\end{equation}}
\newcommand{\eeww}{$e^+ e^- \to W^+ W^-$}

\begin{document}


\preprint{
May 1999
\hspace{-30.5mm}
\raisebox{2.8ex}{SNUTP 99--026}
}

\title{ 
Polarization effects on W boson pair productions \\
with the extra neutral gauge boson 
at the $e^+ e^-$ Linear Collider
}

\author{
Dong-Won Jung$^a$\thanks{dwjung@zoo.snu.ac.kr},
Kang Young Lee$^b$\thanks{kylee@ctp.snu.ac.kr},
H. S. Song$^{a,b}$\thanks{hssong@physs.snu.ac.kr}, 
and Chaehyun Yu$^a$\thanks{chyu@zoo.snu.ac.kr}
}
\vspace{1.0cm}

\address{
Department of Physics, Seoul National University, Seoul 151--742, Korea\\
Center for Theoretical Physics, Seoul National University,
Seoul 151--742, Korea}

\maketitle

\begin{abstract}

We perform the comprehensive analysis of the polarization effects
on the \eeww process in the presence of the extra neutral gauge boson 
at the LC energies.
Consideration of the polarizations of the produced $W$ bosons 
and the beam polarizations 
provides substantial enhancements of the sensitivity
to the $Z$--$Z'$ mixing angles in various models and
the asymmetry variables also give the strict constraints 
on the mixing angles.
We find that the $\chi$--model and the left-right model
get the strict constraint from $\sigma_{LL}^{unpol}$
while the $\psi$--model and the $\eta$--model
from the beam polarization asymmetry.

\end{abstract}

\pacs{ }

\narrowtext

\section{Introduction}

Pair production of $W$ bosons is one of the principal process
to study the electroweak gauge symmetry
in the future $e^+ e^-$ linear colliders (LC).
The cross section for $e^+ e^- \to W^+ W^-$ and its angular distribution
depend upon various properties of the $W$ boson such as mass,
decay widths etc..
This process is also sensitive to the triple gauge boson couplings (TGC)
of the $WW\gamma$ and $WWZ$ vertices
which enable us to study the nonabelian nature of the gauge structure
for the electroweak theory of the Standard Model (SM).
These analyses can be carried out at the  LC energies
($\sqrt{s} \ge 500$ GeV),
promising sensitivities of order 1 $\%$ and expected better \cite{LC}.

Polarizations of the electron and positron beams provide 
very effective tools to investigate the new physics effects,
particularly useful for the \eeww process.
Using the right-handed polarized electron beam,
the $t$-channel neutrino exchange diagram of the SM depicted in Fig. 1 (a)
is switched off, which occupies a large fraction of the events.
The absence of the SM $t$-channel contribution results in 
the sensitivity to the existence of remaining new physics effects 
in the charged current sector and 
the relative enhancements of them in the neutral current interactions
by the great reduction of the SM background.
Furthermore each helicity channel
shows peculiar behavior depending upon the beam polarizations.
Therefore it would be fruitful in search of the new physics 
beyond the SM to consider the polarization
observables together with beam polarizations at this process.

In many scenarios that the gauge symmetries of the SM are extended 
by adding extra symmetries or embedded into a larger gauge group,
we have one or more new heavy neutral gauge bosons.
The recent bounds of the new gauge boson masses come from
the direct search at $p \bar{p}$ collider via Drell--Yan production
and subsequent decay to charged leptons \cite{CDF}
while indirect constraints for the $Z'$ mass and
mixing angles are given from high precision LEP data at $Z$ peak
energy and various low energy neutral current experiment data
\cite{altarelli,langacker,erler,babu,hagiwara,chay}.
In the present work, we consider 
\bea
e^-(k_1,\kappa) + e^+(k_2,-\kappa) 
\to W^-(p_1,\lambda_1) + W^+(p_2,\lambda_2)
\eea
with an $Z'$ boson
involved in SO(10) and string inspired E$_6$ 
grand unified theories (GUTs),
which are theoretically well--motivated and
have been extensively studied in both 
theoretical and experimental fields \cite{hewett,LR}.
Here, $k_i$ and $p_i$ denote the particle momenta 
while $\kappa$ and $\lambda$ the helicities.
Incorporating the extra neutral gauge boson, deviations
of the TGCs from the SM predictions would be also introduced
as well as the corrections to the neutral current interactions.
Identifying the final states as ordinary $W$ bosons,
the TGCs come from the $W_1 W_2 W_3$ term 
out of the SU(2)$_L$ gauge kinetic terms
in the gauge eigenstates
and this term leads to 
an additional $ \xi~WWZ'$ term 
as well as the ordinary $WW\gamma$ and $WWZ$ couplings 
in the presence of the $Z'$ boson.
We note that the additional $WWZ'$ term
always accompanies the $Z$--$Z'$ mixing angle $\xi$.
Thus the $Z'$ exchange diagram depicted in Fig. 1 (b)
is doubly suppressed by the mixing angle $\xi$
and the inverse square of the $Z'$ mass, both of which are expected
to be the same order of magnitudes away from the $Z'$ peak.
Hence the \eeww process with the $Z'$ boson depends
only on $\xi$ keeping the leading corrections
of the $Z'$ boson effects off the $Z'$ resonance.
It is favorable to set the bounds on the mixing angle 
$\xi$ from this process with less contaminations
of other model--dependent parameters
while the $e^+ e^- \to f \bar{f}$ processes
depend on both the mixing angle $\xi$ and the mass of the
$Z'$ boson at the LC.

In this paper, we present the comprehensive analysis of the
polarization effects on the $W$ pair production at the $e^+ e^-$ collider 
with an $Z'$ boson at $\sqrt{s} = 500$ GeV and 1 TeV, 
which are typical center-of-mass (CM) energies of the LC \cite{LC}.
Similar analyses on this topic have been done previously
in the case that the $Z'$ mass is about a few hundred GeV 
\cite{najima,kalyniak},
that the incident beams are polarized \cite{comas},
and that one focuses on the backward direction enhancements \cite{pankov}.
Our analysis is performed with the polarized cross sections 
of the produced $W$ pair and the asymmetry variables 
with or without beam polarizations. 
We assume that the mass of the $Z'$ boson is quite large
and the $Z'$ effects contribute to the process
through only the mixing angle $\xi$.
This paper is organized as follows.
In Section II, we briefly review the models involving 
the additional $Z'$ bosons:
the extra U(1) symmetries and the left-right (LR) model. 
We calculate the helicity amplitudes 
and discuss the possible enhancements of the $Z'$ effects
for the polarized cross sections in Section III.
More analyses on the asymmetry variables are given in Section IV.
Finally we conclude in Section V.

\section{The Models}

When the unifying gauge group breaks to its subgroups,
extra U(1) often appears in many GUT models
as an intermediate stage involving an additional neutral gauge boson.
Here we consider the $\chi$, $\psi$ and $\eta$ models
which occur in the symmetry breaking of SO(10) or string inspired
E$_6$ GUTs according to the path of the gauge symmetry breaking pattern: 
SO(10) $\to$ SU(5)$\times$U(1)$_\chi$,
E$_6$ $\to$ SO(10)$\times$U(1)$_\psi$,
E$_6$ $\to$ rank 5 groups, respectively.
In the latter case, the corresponding $Z'_\eta$ boson is given by 
a linear combination of the $Z'_\chi$ and $Z'_\psi$ such that
$Z'_\eta = \sqrt{3/8} Z'_\chi - \sqrt{5/8} Z'_\psi$.
The LR model based on the SU(2)$_L \times$SU(2)$_R \times$U(1)
gauge group is also considered, which is another subgroup of SO(10).

In the presence of extra neutral gauge bosons,
the relevant terms of the most general lagrangian is given by
\be
{\cal L} = {\cal L}_m + {\cal L}_{NC} + {\cal L}_{CC},
\ee
where  
\bea
{\cal L}_m &=& \frac{1}{2} m_{Z_1}^2 Z_{1 \mu} Z_1^\mu
             + \frac{1}{2} m_{Z_2}^2 Z_{2 \mu} Z_2^\mu
             + \delta m^2 Z_{1 \mu} Z_2^\mu,
\nonumber \\
{\cal L}_{NC} &=& -\frac{1}{2} g_1 \sum_f
                  \bar{f} \gamma_\mu (g^f_{0V} - g^f_{0A} \gamma_5) f Z_1^\mu
                  -\frac{1}{2} g_2 \sum_f
                  \bar{f} \gamma_\mu (h^f_{0V} - h^f_{0A} \gamma_5) f Z_2^\mu.
\eea
In the case of an additional U(1), 
the charged current interaction term ${\cal L}_{CC}$ is same
as that of the SM.
The model parameter $h^f_{0V}$ and $h^f_{0A}$ are listed in the Table I
and the coupling constant of the $Z_2$ boson is given by
\be
g_2 = \sqrt{\frac{5}{3}}~g_1 \sin \theta_W \sqrt{\lambda_g},
\ee
where $\lambda_g$ depends on the symmetry breaking pattern 
and of ${\cal O}(1)$ \cite{hewett}.
We let $\lambda_g =1$ assuming that the gauge group 
breaks directly to SU(3)$_C\times$SU(2)$_L \times$U(1)$_Y$.

\vskip 1.5cm
\noindent
\begin{center}
\begin{tabular}{|c|c|c|c|}
\hline
 &~~~~ $\chi$ model~~~~ &~~~~ $\psi$ model~~~~ &~~~~ $\eta$ model~~~~  \\
\hline
~~~~$h_{V0}$~~~~&~~$2/\sqrt{10}$~~&~~ $ 0$~~ &~ $3/2 \sqrt{15}$~\\
~~~~$h_{A0}$~~~~&~~$1/\sqrt{10}$~~&~~~$ 1/\sqrt{6}$~~~&~$-1/2 \sqrt{15}$~\\
\hline
\end{tabular}
\end{center}
\vspace{0.8cm}
{Table~1}. {\it 
The vector and axial-vector couplings of charged leptons 
to the $Z'$ gauge boson in the SO(10) and $E_6$ GUT models.
}
\vspace{0.8cm}
\vskip 0.5cm

After diagonalizing the mass matrix, we define the mass eigenstates
of neutral gauge bosons as
\be
\left( \begin{array}{c}
       Z_1   \\
       Z_2 
       \end{array}   \right) 
= \left( \begin{array}{cc}
       \cos\xi& -\sin\xi   \\
       \sin\xi&  \cos\xi
       \end{array}   \right) 
\left( \begin{array}{c}
       Z   \\
       Z' 
       \end{array}   \right), 
\ee
where the mixing angle $\xi$ is given by
\be
\tan 2 \xi = -\frac{2 \delta m^2}
                   {m_{Z_2}^2-m_{Z_1}^2}.
\ee
In terms of $Z$ and $Z'$ states, the neutral current interaction
terms are written as
\be
{\cal L}_{NC} = -\frac{1}{2} g_1 \sum_f
                  \bar{f} \gamma_\mu (g^f_V - g^f_A \gamma_5) f Z^\mu
                  -\frac{1}{2} g_2 \sum_f
                  \bar{f} \gamma_\mu (h^f_V - h^f_A \gamma_5) f Z'^\mu,
\ee
with the vector and axial-vector couplings
\be
g^f_{V,A} = g^f_{0V,A} + \xi~ h^f_{0V,A} \frac{g_2}{g_1},~~~
h^f_{V,A} = h^f_{0V,A} - \xi~ g^f_{0V,A} \frac{g_1}{g_2},
\ee
by keeping the leading order in $\xi$.
Since there are no corrections to the charged current sector in this case,
the new physics effects reside in the mixing angle $\xi$ only.

In the case of the LR model,
one can find the detailed review of the LR model lagrangian
in Ref. \cite{chay,LR} and references therein.
Here, we follow the formulae of Ref. \cite{chay}. 
The mass matrix of neutral gauge bosons 
$(B_\mu,W^L_{3 \mu},W^R_{3 \mu})$ is diagonalised
by 3 angles $\theta_W$, $\theta_R$, and $\xi$
to obtain the physical eigenstates $(A_\mu,Z_\mu,Z'_\mu)$.
Among these angles, $\theta_W$ is corresponding to the
Weinberg angle of the SM which describes the $A$--$Z_1$ mixing
while $\xi$ is the $Z$--$Z'$ mixing angle
corresponding to Eq. (6).
Meanwhile $\theta_R$ is an additional parameter depending upon the model.
It is preferred to write the lagrangian in terms of the
left--right basis instead of the form of Eq. (3). 
After the diagonalization, we write the neutral current sector
\begin{eqnarray}
{\cal L}_{\mathrm{NC}} &=& -e \overline{f} A\hspace{-2.5mm}/ \hspace{0.3mm}
\Bigl[ (T_{L3} +S)
P_L + (T_{R3} +S) P_R \Bigr] f \nonumber \\
&&- \overline{f} Z\hspace{-2.7mm}/\hspace{0.48mm}
\Bigl[ \Bigl( g_L c_W c_{\xi} T_{L3} -
g_1 (c_R s_W c_{\xi} + s_R s_{\xi} ) S\Bigr) P_L \nonumber \\
&&\ \ + \Bigl( g_R (c_R s_{\xi} - s_R s_W c_{\xi}) T_{R3} - g_1 (c_R
s_W c_{\xi} + s_R s_{\xi} )S \Bigr) P_R \Bigr] f \nonumber \\
&&- \overline{f} Z^{\prime}\hspace{-3.6mm}/ \hspace{1.3mm}
\Bigl[ \Bigl(-g_L c_W
s_{\xi} T_{L3} + g_1 (c_R s_W s_{\xi} - s_R c_{\xi})S \Bigr) P_L
\nonumber \\
&&\ \ + \Bigl( g_R (c_R c_{\xi} + s_R s_W s_{\xi} ) T_{R3} + g_1 (c_R
s_W  s_{\xi} - s_R c_{\xi})S \Bigr) P_R \Bigr] f,
\end{eqnarray}
where $c_{i} = \cos \theta_i$, $s_{i} = \sin \theta_i$ and
$c_{\xi} = \cos \xi$, $s_{\xi} = \sin \xi$.
Here $g_L$ is the gauge coupling constant of SU(2)$_L$ group,
$g_R$ is that of SU(2)$_R$, $g_1$ is that of U(1) 
and $S$ is the U(1) charge.
Since we consider the general case of the model, 
$g_L$ need not be same as $g_R$.
Looking at the couplings to the photon,
we define the electric charge $Q = T_{L3}+T_{R3}+S$
and obtain the relations among the couplings:
\be
e = g_L s_W = g_R s_R c_W = g_1 c_R c_W.
\ee
The neutral current coupled to the $Z$ boson 
up to the leading order in $\xi$ is given by
\be
J_Z^{\mu} = \frac{e}{s_W c_W} \overline{f} \gamma^{\mu} \Bigl
[ T_{L3} - s_W^2 Q + \xi s_W \Bigl( t_R (T_{L3} -Q) + (t_R +
\frac{1}{t_R}) T_{R3} \Bigr)\Bigr] f,
\label{zcurrent}
\ee
where $t_R = s_R/c_R$.

In the LR model, there is also a mixing in the charged current sector
which affects the $t$--channel diagram.
However if we fix the final states as the ordinary $W$ boson pair,
both of the charged current interactions should be
right-handed and the exchanged neutrino is also the
right-handed one.
As a result, the correction to the $t$--channel diagram is
suppressed by the quadratic factor of the $W$ boson mixing angle 
and the inverse of the heavy right-handed neutrino mass
even in the leading order of corrections.
It is natural to regard the mixing angle of the charged current sector 
as the same order of magnitude of that of the neutral current sector.
Hence we can safely neglect the corrections of the LR model 
to the $t$--channel when keeping the leading corrections of 
extra gauge boson effects.
We conclude that we can concentrate on the correction to the
$Z e^+ e^-$ vertex for the $Z'$ boson effects even in the LR model.

\section{$W$ boson pair production}

The $W$ pair is produced through the $s$--channel diagrams
mediated by neutral gauge bosons and the $t$--channel mediated
by neutrino shown in Fig.1.
Denoting the helicity of the electron by
$\kappa = \pm 1$ and the helicities of the $W^-$ and $W^+$ bosons
by $\lambda_1$ and $\lambda_2$, respectively as in the Eq. (1),
the helicity amplitudes are given by 
\begin{equation}
{\cal M}_{SM}^\kappa(\lambda_1,\lambda_2)=\frac{e^2}{2\sin^2\theta_W}
         \frac{1}{t}
        {\cal M}_1^\kappa \delta_{\kappa -} + e^2\Big(\frac{1}{s}-
	 \frac{\epsilon_\kappa }{\sin^2\theta_W} \frac{1}{s-m_Z^2} \Big)
        (2 {\cal M}_3^\kappa - {\cal M}_2^\kappa),
\end{equation}
where the $Z e^+ e^-$ coupling is
\begin{equation}
\epsilon_\kappa = - \frac{1}{2} \delta_{\kappa -} + \sin^2\theta_W.
\end{equation}
The ${\cal M}_i^\kappa$ are given by
\begin{eqnarray}
{\cal M}_1^\kappa &=& \overline{v}(p_2) / \hspace{-1.8mm}\epsilon_2^\ast
   ( / \hspace{-1.8mm}k_2- / \hspace{-1.8mm}p_2) / 
       \hspace{-1.8mm}\epsilon_1^\ast
   \omega_\kappa u(p_1), 
\nonumber \\
{\cal M}_2^\kappa &=& \overline{v}(p_2) 
           ( / \hspace{-1.8mm}k_2 - / \hspace{-1.8mm}k_1)
           (\epsilon_1^\ast \cdot\epsilon_2^\ast) \omega_\kappa u(p_1), 
\nonumber \\
{\cal M}_3^\kappa &=& \overline{v}(p_2)[ / \hspace{-1.8mm} \epsilon_2^\ast
   (\epsilon_1^\ast\cdot k_2)- / \hspace{-1.8mm}\epsilon_1^\ast
   (\epsilon_2^\ast \cdot k_1)] \omega_\kappa u(p_1) ,
\end{eqnarray}
where $\omega_\kappa$ is the helicity projection operator 
and $\epsilon_1^\ast$ $(\epsilon_2^\ast)$ is a polarization vector 
of $W^-(W^+)$.  

Keeping only the leading corrections in $\xi$,
the scattering amplitude is written as
\begin{eqnarray}
{\cal M}^\kappa (\lambda_1,\lambda_2) 
&=& {\cal M}_{SM}^\kappa (\lambda_1,\lambda_2) 
   +\xi{\cal M}_{NP}^\kappa (\lambda_1, \lambda_2), 
\nonumber \\
{\cal M}_{NP}^\kappa (\lambda_1,\lambda_2) 
&=& \frac{e^2}{\sin^2\theta_W}\frac{1}{s-m_Z^2}
       (h_{0V}^e - \kappa h_{0A}^e)
       (2 {\cal M}_3^\kappa - {\cal M}_2^\kappa), 
\end{eqnarray}
where $h_{0V}^e$ and $ h_{0A}^e$ are model-dependent parameters
given in the previous section.
Hereafter we drop the superscript $e$.

In the CM frame, the helicity amplitudes are given by
\bea
&{\cal M}^-(+,+) &= e^2 \sin \theta \left[
\frac{1}{\sin^2\theta_W} \frac{s}{4t} (\cos \theta - \beta) 
- \alpha_L \beta 
\right],
\nonumber \\
&{\cal M}^-(+,0) &= e^2 \frac{\sqrt{2}}{\sqrt{1-\beta^2}}(\cos \theta+1) 
\left[
\frac{1}{2\sin^2\theta_W} \frac{s}{4t} (2 \beta+1-2\cos\theta-\beta^2) 
+ \alpha_L \beta 
\right],
\nonumber \\
&{\cal M}^-(+,-) &= e^2 \sin \theta (\cos\theta+1) 
\frac{1}{\sin^2\theta_W} \frac{s}{4t},
\nonumber \\
&{\cal M}^-(0,+) &= e^2 \frac{\sqrt{2}}{\sqrt{1-\beta^2}}(\cos \theta-1) 
\left[
\frac{1}{2\sin^2\theta_W} \frac{s}{4t} (2 \beta-1-2\cos\theta+\beta^2) 
+ \alpha_L \beta 
\right],
\nonumber \\
&{\cal M}^-(0,0) &= e^2 \frac{1}{1-\beta^2}\sin \theta 
\left[
\frac{1}{4\sin^2\theta_W} \frac{s}{4t} (3 \beta-2\cos\theta-\beta^3) 
+ \alpha_L \beta  (3-\beta)
\right],
\nonumber \\
&{\cal M}^-(-,+) &= e^2 \sin \theta (\cos\theta-1) 
\frac{1}{\sin^2\theta_W} \frac{s}{4t},
\eea
and
\bea
&{\cal M}^+(+,+) &= e^2 \alpha_R \beta  \sin \theta,
\nonumber \\
&{\cal M}^+(+,0) &= e^2 \frac{\sqrt{2}}{\sqrt{1-\beta^2}}
\alpha_R \beta \sin \theta (\cos \theta-1), 
\nonumber \\
&{\cal M}^+(+,-) &= 0,
\nonumber \\
&{\cal M}^+(0,+) &= e^2 \frac{\sqrt{2}}{\sqrt{1-\beta^2}}
\alpha_R \beta  \sin \theta (\cos \theta+1), 
\nonumber \\
&{\cal M}^+(0,0) &= e^2 \frac{1}{1-\beta^2}
\alpha_R \beta \sin \theta (3-\beta^2), 
\nonumber \\
&{\cal M}^+(-,+) &= 0,
\eea
where $\beta = \sqrt{1-4m_W^2/s}$, $theta$ is the scattering angle and
\bea
\alpha_L &= \left(
\displaystyle{
1-\frac{s}{s-m_{Z}^2} \frac{1}{2\sin^2\theta_W}(g_V+g_A)} \right),
\nonumber \\
\alpha_R &= \left(
\displaystyle{
1-\frac{s}{s-m_{Z}^2} \frac{1}{2\sin^2\theta_W}(g_V-g_A)} \right),
\eea
with $g_V$ and $g_A$ given in Eq. (8).
The CP invariance leads to
\bea
{\cal M}^\kappa(0,\pm)= {\cal M}^\kappa(\mp,0),~~~~
{\cal M}^\kappa(+,+)= {\cal M}^\kappa(-,-).
\eea
These results agree with the formulae in Ref. \cite{denner}.

The differential cross section is obtained by the sum
of the polarized cross sections
\begin{equation}
\frac{d \sigma_{total}}{d \cos \theta} 
= \frac{1}{4} \sum_{\lambda_1,\lambda_2} 
\left[
  (1+P_-)(1-P_+)\frac{d \sigma^+ (\lambda_1,\lambda_2)}{d \cos \theta}+
  (1-P_-)(1+P_+)\frac{d \sigma^- (\lambda_1,\lambda_2)}{d \cos \theta}
\right],
\end{equation}
where $P_-$ $(P_+)$ is the polarization of the electron 
(positron) beam and the polarized differential cross sections are given by
\begin{equation}
\frac{d \sigma^\kappa (\lambda_1,\lambda_2)}{d \cos \theta}
= \frac{\beta}{32 \pi s} | {\cal M}^\kappa (\lambda_1,\lambda_2)|^2.
\end{equation}

In order to estimate the search bound of the $Z'$ contributions,
we take the linear approximation of the cross section
under the assumption that the mixing angle $\xi$ is very small:
\begin{equation}
\sigma = \sigma_{SM} + \xi \sigma_1.
\end{equation}
We plot the ratio of the correction term $\sigma_1$ 
to the SM cross section $\sigma_{SM}$ 
with varying the CM energy $\sqrt{s}$ in Fig. 2 and 3,
when the electron beam is unpolarized 
and right-handed polarized respectively.
When the unpolarized electron beam is used,
the large  neutrino exchange $t$--channel contribution
conceals the new physics effects.
Considering the cross section of both longitudinally polarized
$W$ boson pair, however, we find that 
$|\sigma_1^{LL}/\sigma_{SM}^{LL}|$ increases 
according to the increase of $\sqrt{s}$ in the Fig. 2,
which implies that the observable $\sigma_{LL}$ 
becomes sensitive to the $Z'$ corrections 
at high energy collisions.
In the limit of $\sqrt{s} \gg m_{_Z}$,
the leading term of the SM helicity amplitude of 
both longitudinally polarized $W$ pair,
${\cal M}^\kappa_{SM} (0,0)$ is of order of $m_{_Z}^2/s$ 
while that of ${\cal M}^\kappa_{NP} (0,0)$ is of order 1.
As a result, the ratio is given by 
\begin{equation}
\left|
\frac{\sigma_1^{LL}}{\sigma_{SM}^{LL}} 
\right|
\sim 
\frac{{\cal M}_{SM}(0,0) {\cal M}_{NP}(0,0)}
     {{{\cal M}_{SM}(0,0)}^2}
\propto \frac{s}{m_{_Z}^2},
\end{equation}
which rapidly increases along with the CM energy
and results in the sensitivity to the $Z'$ effects.
We expect to probe the mixing angle $\xi$ more precisely
from the observable $\sigma^{LL}$ than from the total cross section.

Even though the polarized cross section $\sigma^{LL}$
is sensitive to the mixing angle $\xi$, 
we should put up with a statistical loss
because $\sigma^{LL}$ is much smaller than $\sigma_{tot}$.
Alternatively we suggest to use the right-handed electron beam
which avoids masking the large $t$--channel contribution
and also show the remarkable feature like Eq. (23).
All of the SM helicity amplitudes for right-handed electron beam,
${\cal M}^+_{SM}$, are proportional to the factor
\begin{equation}
\frac{1}{s} - \frac{1}{s-m_{_Z}^2} 
= -\frac{m_{_Z}^2}{s} \cdot \frac{1}{s-m_{_Z}^2},
\end{equation}
while ${\cal M}_{NP}^+$ are proportional to $1/(s-m_{_Z}^2)$
with the common part of the helicity amplitudes
$(2 {\cal M}^+_3 - {\cal M}^+_2)$ in the Eq. (15).
Accordingly all ratios of helicity cross sections $|\sigma_1/\sigma_{SM}|$ 
are always proportional to $s/m_{_Z}^2$
as shown in the Fig. 3.
Thus the sensitivity to the new physics effects
is yielded even for the total cross section 
if we use the right-handed electron beam. 
Moreover it would be statistically favorable 
to use the total cross section with the right-handed polarized 
electron beam than to use $\sigma_{LL}$ with unpolarized beam.

\vskip 1.7cm
\noindent
\begin{center}
\begin{tabular}{|c|rr|rr|}
\hline
&\multicolumn{2}{c|}{$\sigma_{LL}^{unpol}$}
&\multicolumn{2}{c|}{$\sigma_{tot}^{90\%R}$}\\
\cline{2-5}
\raisebox{\baselineskip}%
[0cm][0cm]{}
&\multicolumn{1}{c}{$\sqrt{s}=$500 GeV}
&\multicolumn{1}{c|}{1 TeV}
&\multicolumn{1}{c}{500 GeV}
&\multicolumn{1}{c|}{1 TeV}\\ \hline
~~~~$\chi$~~~~
&$\begin{array}{r}-0.00139\\0.00135\end{array}$
&$\begin{array}{r}-0.00034\\0.00033\end{array}$
&$\begin{array}{r}-0.00604\\0.00557\end{array}$
&$\begin{array}{r}-0.00175\\0.00159\end{array}$\\ \hline
$\psi$
&$\begin{array}{r}-0.00782\\0.00635\end{array}$
&$\begin{array}{r}-0.00186\\0.00154\end{array}$
&$\begin{array}{r}-0.00654\\0.00753\end{array}$
&$\begin{array}{r}-0.00184\\0.00216\end{array}$\\ \hline
$\eta$
&$\begin{array}{r}-0.00306\\0.00295\end{array}$
&$\begin{array}{r}-0.00076\\0.00074\end{array}$
&$\begin{array}{r}-0.00479\\0.00437\end{array}$
&$\begin{array}{r}-0.00138\\0.00124\end{array}$\\ \hline
LR
&$\begin{array}{r}-0.00111\\0.00108\end{array}$
&$\begin{array}{r}-0.00027\\0.00027\end{array}$
&$\begin{array}{r}-0.00373\\0.00345\end{array}$
&$\begin{array}{r}-0.00108\\0.00098\end{array}$\\ \hline
\end{tabular}
\end{center}
\vspace{0.8cm}
{Table~2}. {\it 
The reaches of the bounds of the mixing angle $\xi$ from the cross section
for the both longitudinally polarized $W$ pair with unpolarized
$e^-$ beam and the total cross section with 90$\%$ right-handed polarized
$e^-$ beam. The angular cut $|\cos\theta| < 0.9$ is applied.
}
\vspace{0.8cm}

Considering the statistical error and $1 \%$ systematic error, 
the expected bounds of the mixing angle $\xi$
are derived from the polarization observables discussed above
at 95$\%$ C.L. with the angular cut $|\cos \theta| < 0.9$
and presented in Table II.
The integrated luminosities are taken to be as
$\int {\cal L}=50$ fb$^{-1}$ for $\sqrt{s} = 500$ GeV
and $\int {\cal L}=200$ fb$^{-1}$ for $\sqrt{s} = 1$ TeV \cite{LC}.
Note that the generation of 100$\%$ polarization of the electron
beam is hardly reached in practice.
We consider the expected polarization of the $e^-$ beam as 90$\%$ here.
In this case, it still makes a contamination from the 
neutrino exchange $t$--channel of the SM contribution,
especially in the forward scattering region.
The angular cut may reduce the contamination from the $t$--channel,
which should be introduced by the practical reason of 
the detector geometry.
We find that more strict bounds on $\xi$ are derived
from $\sigma^{unpol}_{LL}$ in most cases.
For the illustrative purpose we plot the 
$\sigma^{unpol}_{LL}$ and
$\sigma^{90\% R}_{tot}$
with respect to the mixing angle $\xi$ in the Fig. 4 and 5.

\section{Asymmetries}

Parity violation of the electroweak theory is implied by
the asymmetry between the left and right couplings of the weak 
neutral current interaction.
The asymmetry variables are essential touchstones 
to explore the gauge structure of the SM.
Labelling the cross sections of the \eeww process 
by the direction of $W^\pm$ bosons
and the helicity of the incident electron beam,
we define the forward-backward asymmetry 
\bea
A_{FB} = \frac{\sigma_F - \sigma_B} 
              {\sigma_F + \sigma_B},
\eea
where
\bea
\sigma_F = \int^1_0 d \cos \theta \frac{d \sigma}{d \cos \theta},~~~
\sigma_B = \int^0_{-1} d \cos \theta \frac{d \sigma}{d \cos \theta},
\eea
and the beam polarization asymmetry 
\bea
A_{pol} = \frac{\sigma_R - \sigma_L} 
              {\sigma_R + \sigma_L},
\eea
where $\sigma_{L(R)}$ is the cross section with the left- (right-) handed
electron beam.

We show the forward-backward asymmetry for each model in the Fig. 6
with respect to the mixing angle $\xi$ to estimate the bounds of it.
The beam polarization asymmetry can be defined by both the differential
cross sections and integrated ones.
Figure 7 shows the one from the total cross sections 
with respect to $\xi$
while Fig. 8 shows the asymmetries from the differential
cross sections with respect to $\cos \theta$.
The dotted and dashed lines in the Fig. 8 is corresponding to
the most recent bounds on the mixing angle $\xi$ given in 
Ref. \cite{erler,chay}.
By means of the same manner in the previous section, 
we estimate the reaches of the bounds of $\xi$ 
from the asymmetry variables and present the results in Table III.
We consider the $1\%$ systematic and the statistical errors
at $95\%$ C.L..
It is to be noted that the lower bound of $\xi$
from the forward-backward asymmetry should not be considered seriously
since the higher order corrections in $\xi$ become important
as shown in the Fig. 6.

\vskip 1.5cm
\noindent
\begin{center}
\begin{tabular}{|c|rr|rr|}
\hline
&\multicolumn{2}{c|}{$A_{FB}$}
&\multicolumn{2}{c|}{$A_{pol}$}\\
\cline{2-5}
\raisebox{\baselineskip}%
[0cm][0cm]{}
&\multicolumn{1}{c}{$\sqrt{s}=$500 GeV}
&\multicolumn{1}{c|}{1 TeV}
&\multicolumn{1}{c}{500 GeV}
&\multicolumn{1}{c|}{1 TeV}\\ \hline
~~~~$\chi$~~~~
&$\begin{array}{r}-0.00404\\0.00385\end{array}$
&$\begin{array}{r}-0.00110\\0.00104\end{array}$
&$\begin{array}{r}-0.00220\\0.00216\end{array}$
&$\begin{array}{r}-0.00057\\0.00056\end{array}$\\ \hline
$\psi$
&$\begin{array}{r}-0.02225\\0.00450\end{array}$
&$\begin{array}{r}-0.02235\\0.00429\end{array}$
&$\begin{array}{r}-0.00150\\0.00154\end{array}$
&$\begin{array}{r}-0.00039\\0.00041\end{array}$\\ \hline
$\eta$
&$\begin{array}{r}-0.00920\\0.00840\end{array}$
&$\begin{array}{r}-0.00250\\0.00226\end{array}$
&$\begin{array}{r}-0.00126\\0.00123\end{array}$
&$\begin{array}{r}-0.00033\\0.00032\end{array}$\\ \hline
LR
&$\begin{array}{r}-0.00324\\0.00309\end{array}$
&$\begin{array}{r}-0.00088\\0.00083\end{array}$
&$\begin{array}{r}-0.00120\\0.00118\end{array}$
&$\begin{array}{r}-0.00031\\0.00031\end{array}$\\ \hline
\end{tabular}
\end{center}
\vspace{0.8cm}
{Table~3}. {\it 
The reaches of the bounds of the mixing angle $\xi$ 
from the forward-backward asymmetry $A_{FB}$ and 
the beam polarization asymmetry $A_{pol}$.
The angular cut $|\cos\theta| < 0.9$ is applied.
}
\vspace{0.8cm}

\section{Concluding Remarks}

We study the polarization effects on the \eeww process
with an extra neutral gauge boson.
Since the corrections proportional to the mixing angle $\xi$
relatively increase compared with the SM predictions in this process 
as shown in the Fig. 2,
we expect to obtain the strict bounds on $\xi$
at the high energy collisions of the $e^+ e^-$ linear collider.
When the CM energy reach to 1 TeV,
we show that the $Z$--$Z'$ mixing angle could be measured
up to the order of $10^{-4}$.
On the other hand,
we could set the bounds on the mixing angle
with less contaminations from the other parameter $m_{Z'}$
in this process 
contrary to the fermion pair productions
if we keep the leading order in $\xi$.

In our analysis, the $\chi$--model and the left-right model
are shown to get the most strict constraint 
from the cross section for the both longitudinally
polarized $W$ bosons, $\sigma_{LL}^{unpol}$
while the $\psi$--model and the $\eta$--model
get it from the beam polarization asymmetry $A_{pol}$.
We conclude that the consideration of the polarization effects
on $W$ boson pair production
enable us to probe the $Z$--$Z'$ mixing angle more precisely.
Meanwhile, if we can improve the beam polarization
or reduce the systematic errors,
we expect that the total cross section
with the right-handed electron beam
could provide better results since it has a
statistical advantage compared with $\sigma^{unpol}_{LL}$.

\begin{center}
{\bf Acknowledgements}
\end{center}

This work is supported 
in part by the Korean Science and Engineering Foundation (KOSEF) 
through the SRC program of the Center for Theoretical Physics (CTP)
at Seoul National University
and in part by the Korea Research Foundation 
as the program of 1998.

\newpage
{\Large \bf Figure Captions}
\vskip 2cm

\begin{description}

\item
Fig. 1 :
Feynman diagrams for the \eeww process.

\item
Fig. 2 :
The ratios of the correction terms to the Standard Model predictions,
$|\sigma_1/\sigma_{SM}|$ with respect to the CM energy
when we use the unpolarized electron beam.
The solid lines denote the ratios of total cross sections,
the dashed lines the ratios of the cross sections of $W$ pair
with one longitudinally polarized but one transversely polarized,
and the dotted lines the ratios of the cross sections
of both longitudinally polarized $W$ pair.

\item
Fig. 3 :
The ratios of the correction terms to the Standard Model predictions,
$|\sigma_1/\sigma_{SM}|$ with respect to the CM energy
when we use the right-handed electron beam.
The solid lines denote the ratios of total cross sections,
the dashed lines the ratios of the cross sections of $W$ pair
with one longitudinally polarized but one transversely polarized,
and the dotted lines the ratios of the cross sections
of both longitudinally polarized $W$ pair.

\item
Fig. 4 :
The cross sections for the both longitudinally polarized
$W$ bosons $\sigma^{unpol}_{LL}$ 
with respect to the mixing angle $\xi$ 
when we use the unpolarized electron beam.
The dotted lines denote the cross sections of $\chi$ model,
the short--dashed lines those of $\psi$ model,
the long--dashed lines those of $\eta$ model,
and the dash--dotted lines those of LR model.

\item
Fig. 5 :
The total cross sections $\sigma^{90\% R}_{total}$ 
with respect to the mixing angle $\xi$ 
when we use the 90$\%$ right-handed electron beam.
The dotted lines denote the cross sections of $\chi$ model,
the short--dashed lines those of $\psi$ model,
the long--dashed lines those of $\eta$ model,
and the dash--dotted lines those of LR model.

\item
Fig. 6 :
The forward-backward asymmetry $A_{FB}$
with respect to the mixing angle $\xi$. 
The dotted lines denote the asymmetries of $\chi$ model,
the short--dashed lines those of $\psi$ model,
the long--dashed lines those of $\eta$ model,
and the dash--dotted lines those of LR model.

\item
Fig. 7 :
The beam polarization asymmetry $A_{pol}$
with respect to the mixing angle $\xi$. 
The dotted lines denote the asymmetries of $\chi$ model,
the short--dashed lines those of $\psi$ model,
the long--dashed lines those of $\eta$ model,
and the dash--dotted lines those of LR model.

\item
Fig. 8 :
The beam polarization asymmetry $A_{pol}$
from the differential cross sections
with respect to the scattering angle $\cos\theta$. 
The solid lines denote the Standard Model predictions,
while the dashed lines the maximal deviations with
the mixing angle bounds given in Ref. \cite{erler}.

\end{description}

\newpage

\begin{figure}[h]
\vspace{7.0cm}
\centering
\centerline{\epsfig{file=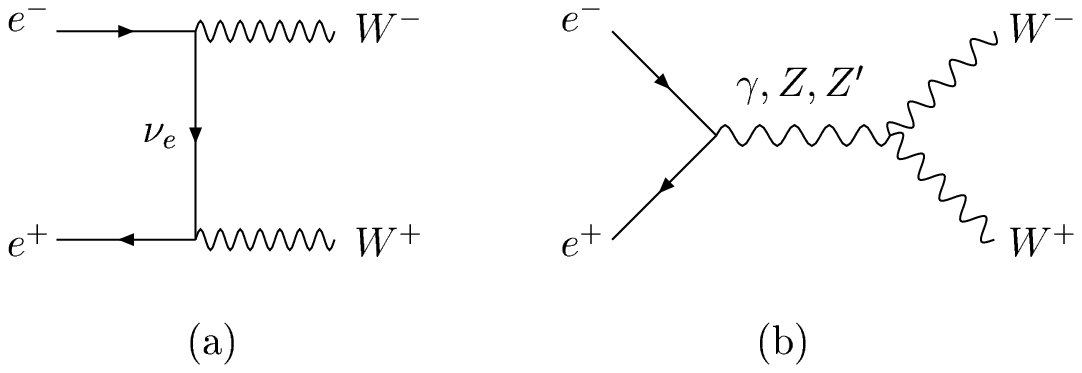}}
\end{figure}
\vspace{2.0cm}
\begin{center}
{\large Fig. 1}
\end{center}

\begin{figure}[th]
\centering
\centerline{\epsfig{file=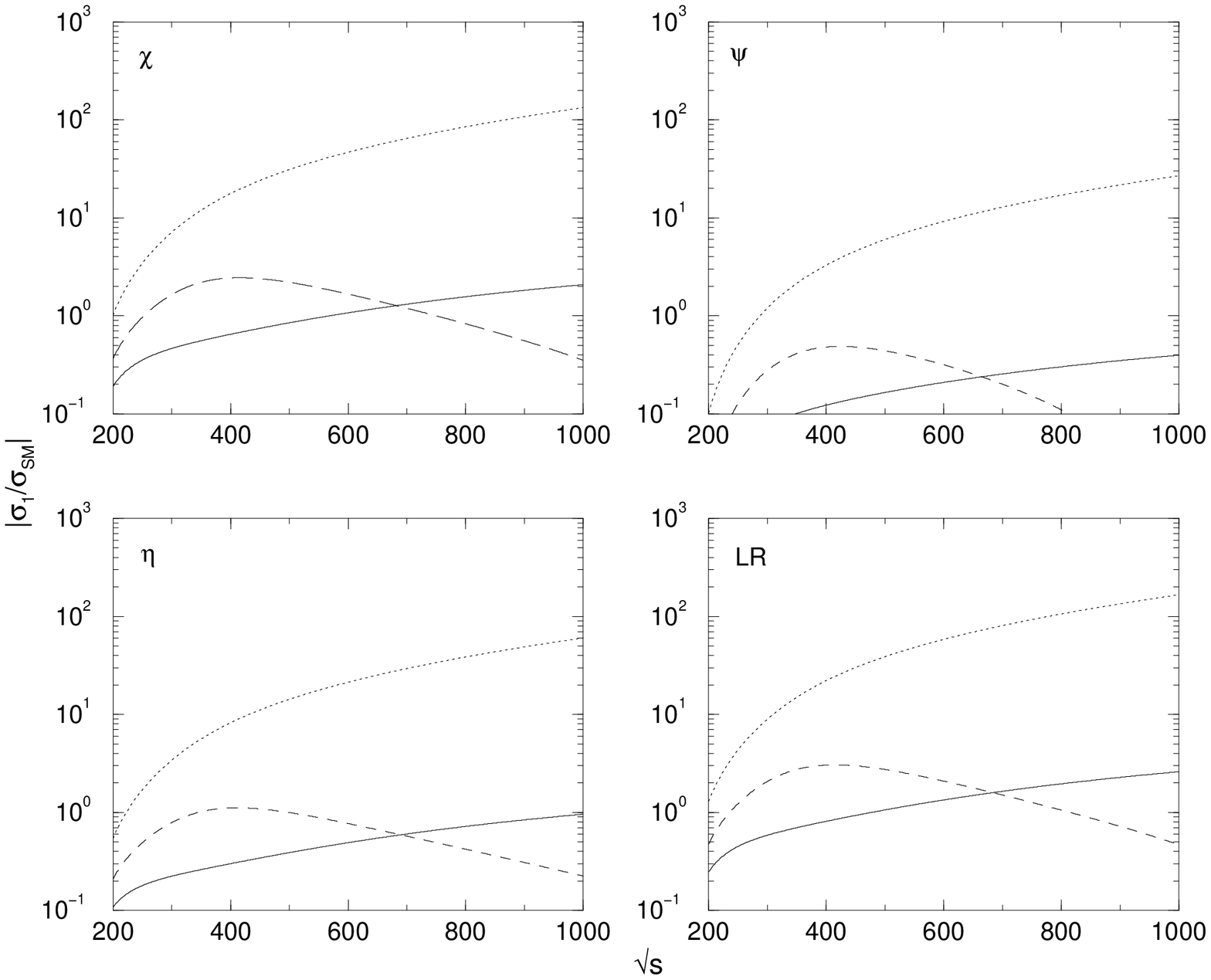}}
\end{figure}
\vspace{1.0cm}
\begin{center}
{\large Fig. 2}
\end{center}

\begin{figure}[th]
\centering
\centerline{\epsfig{file=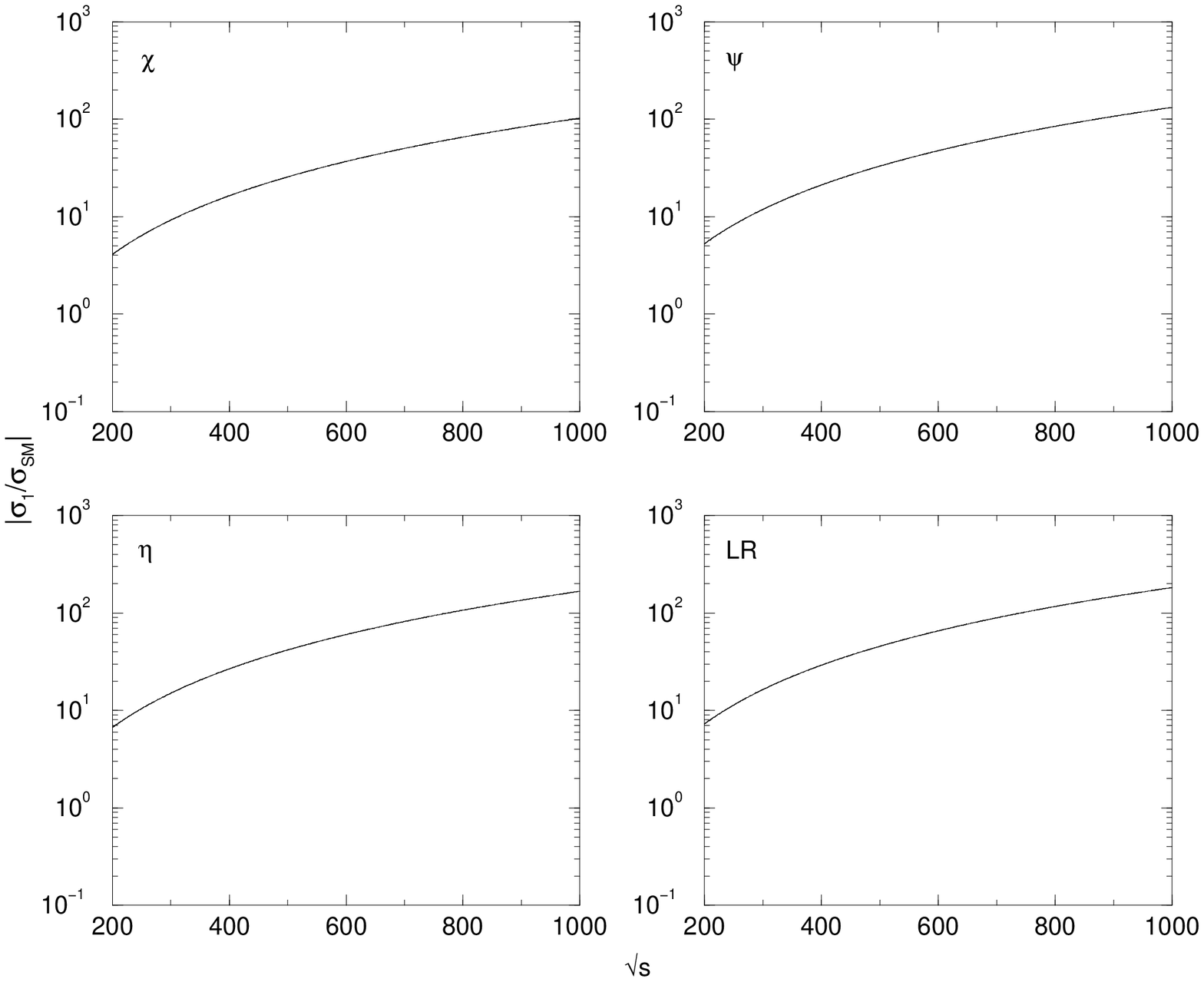}}
\end{figure}
\vspace{1.0cm}
\begin{center}
{\large Fig. 3}
\end{center}

\begin{figure}[th]
\centering
\centerline{\epsfig{file=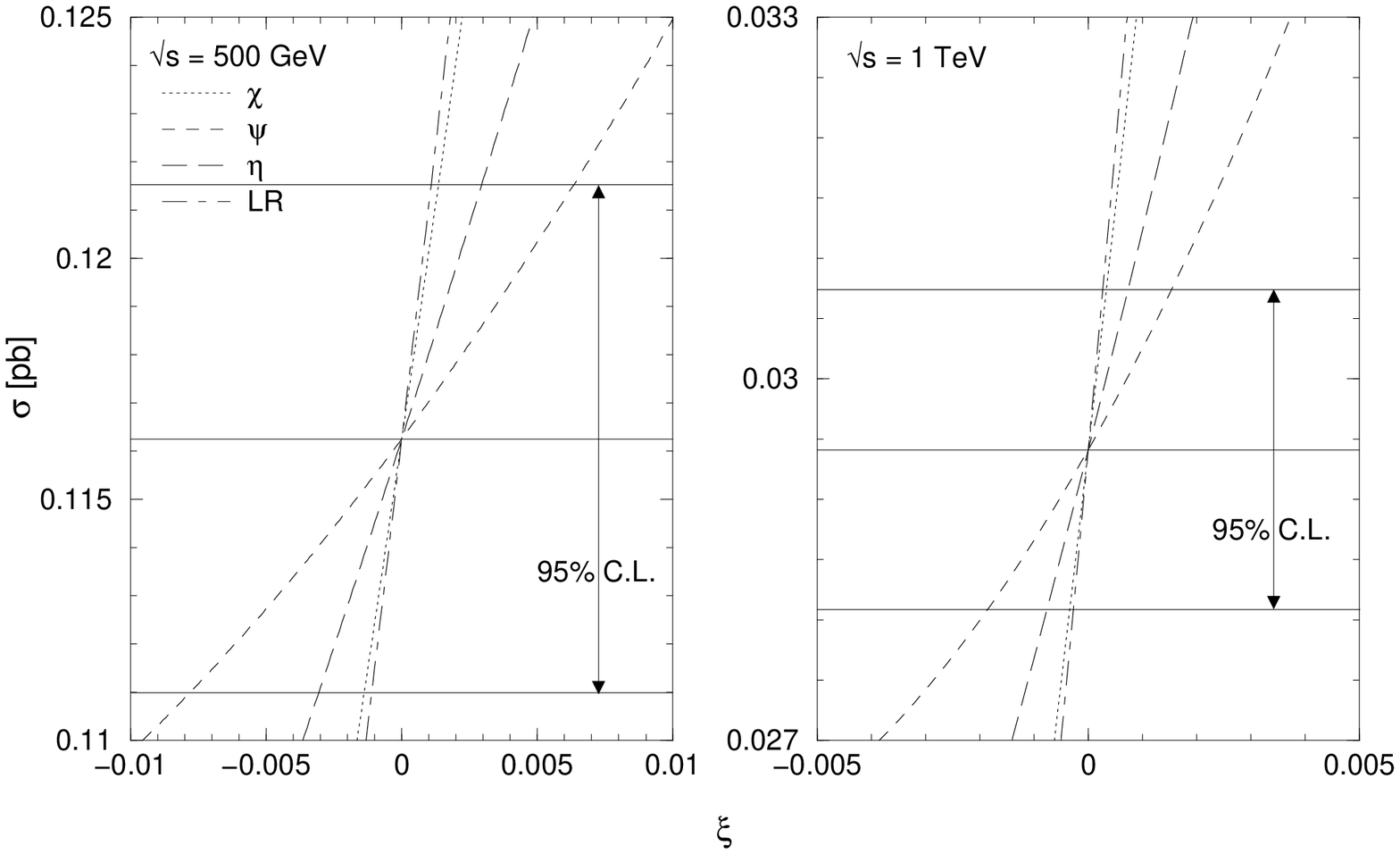,width=19.5cm}}
\end{figure}
\vspace{1.0cm}
\begin{center}
{\large Fig. 4}
\end{center}

\begin{figure}[th]
\centering
\centerline{\epsfig{file=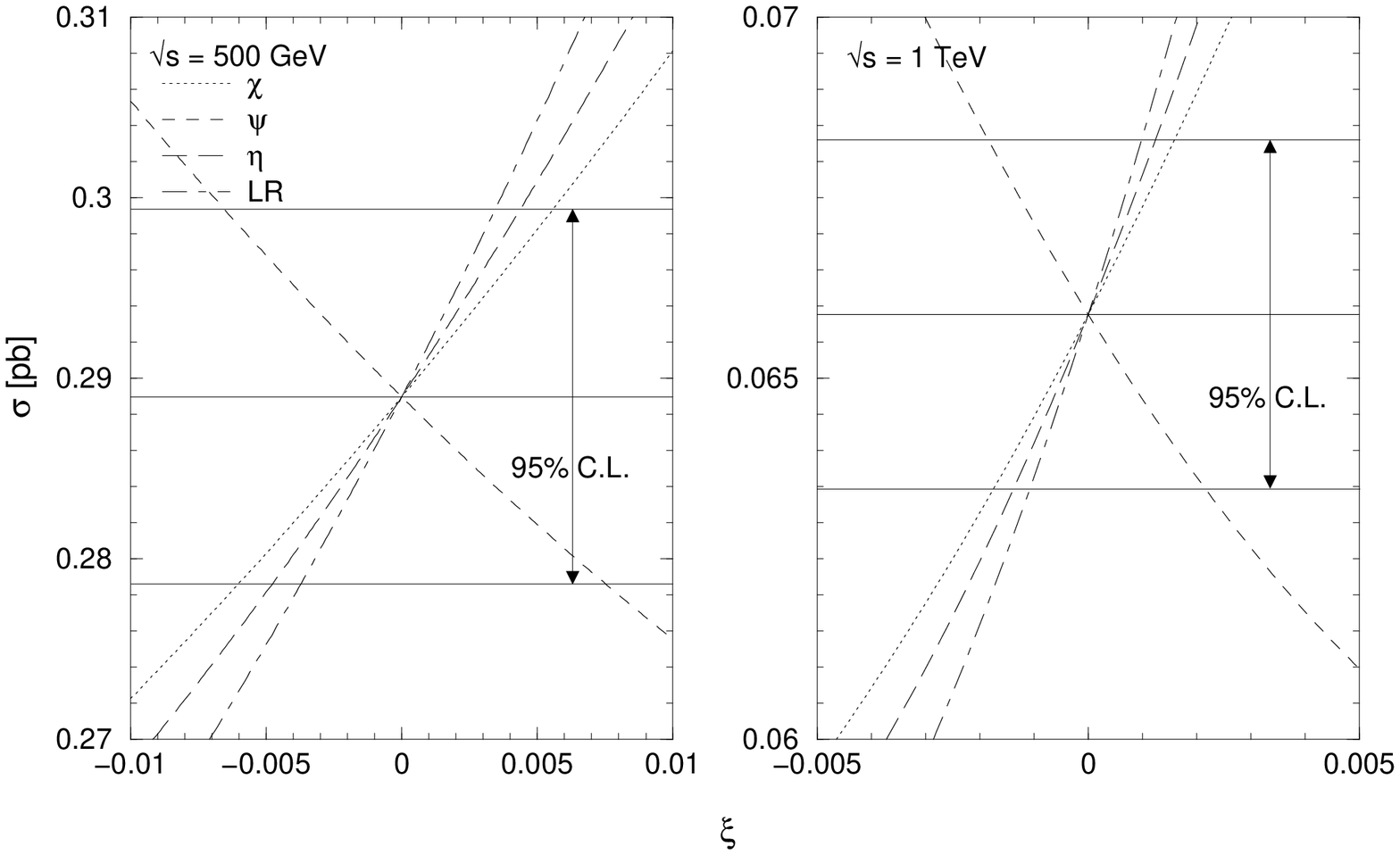,width=19.5cm}}
\end{figure}
\vspace{1.0cm}
\begin{center}
{\large Fig. 5}
\end{center}

\begin{figure}[th]
\centering
\centerline{\epsfig{file=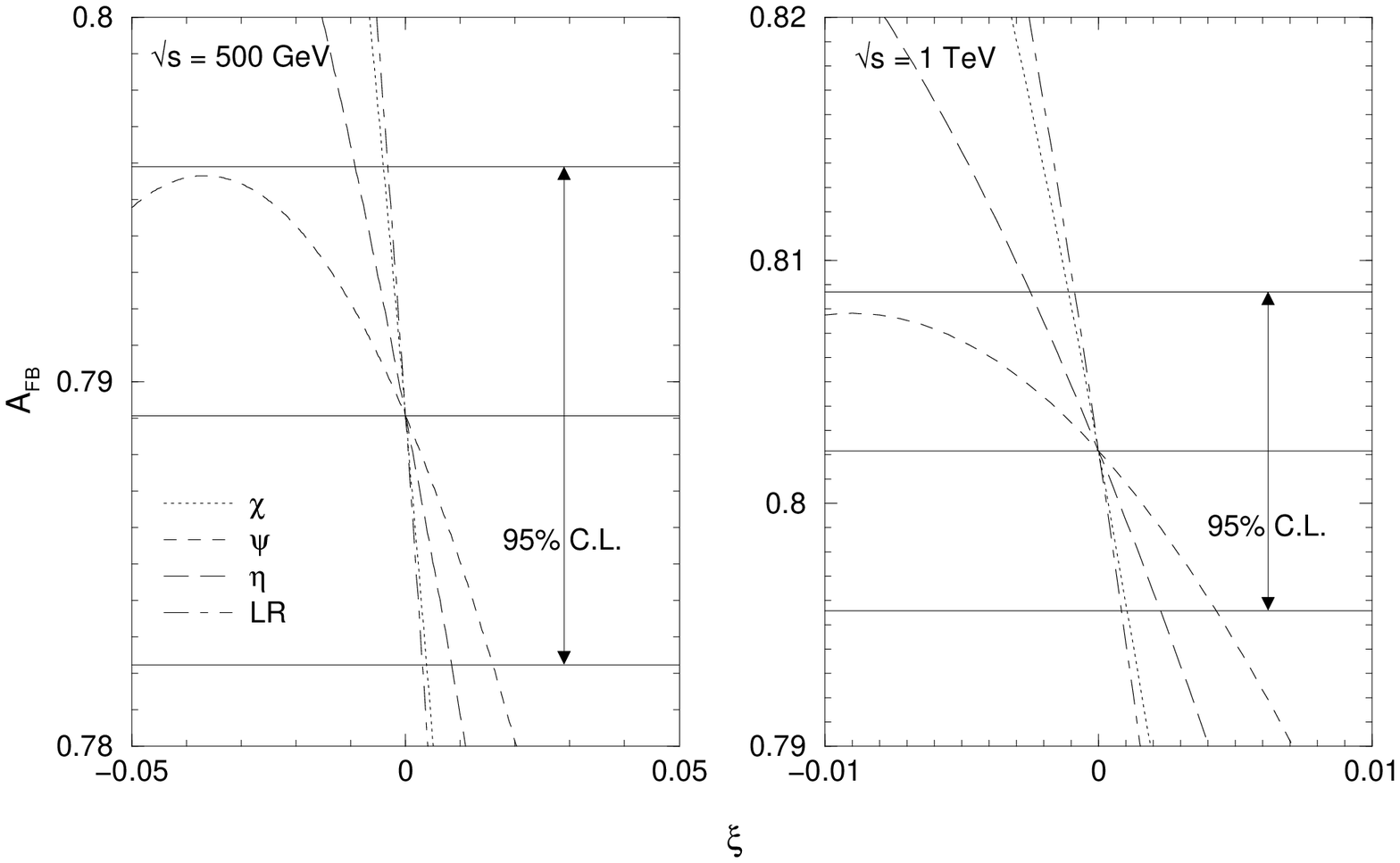}}
\end{figure}
\vspace{1.0cm}
\begin{center}
{\large Fig. 6}
\end{center}

\begin{figure}[th]
\centering
\centerline{\epsfig{file=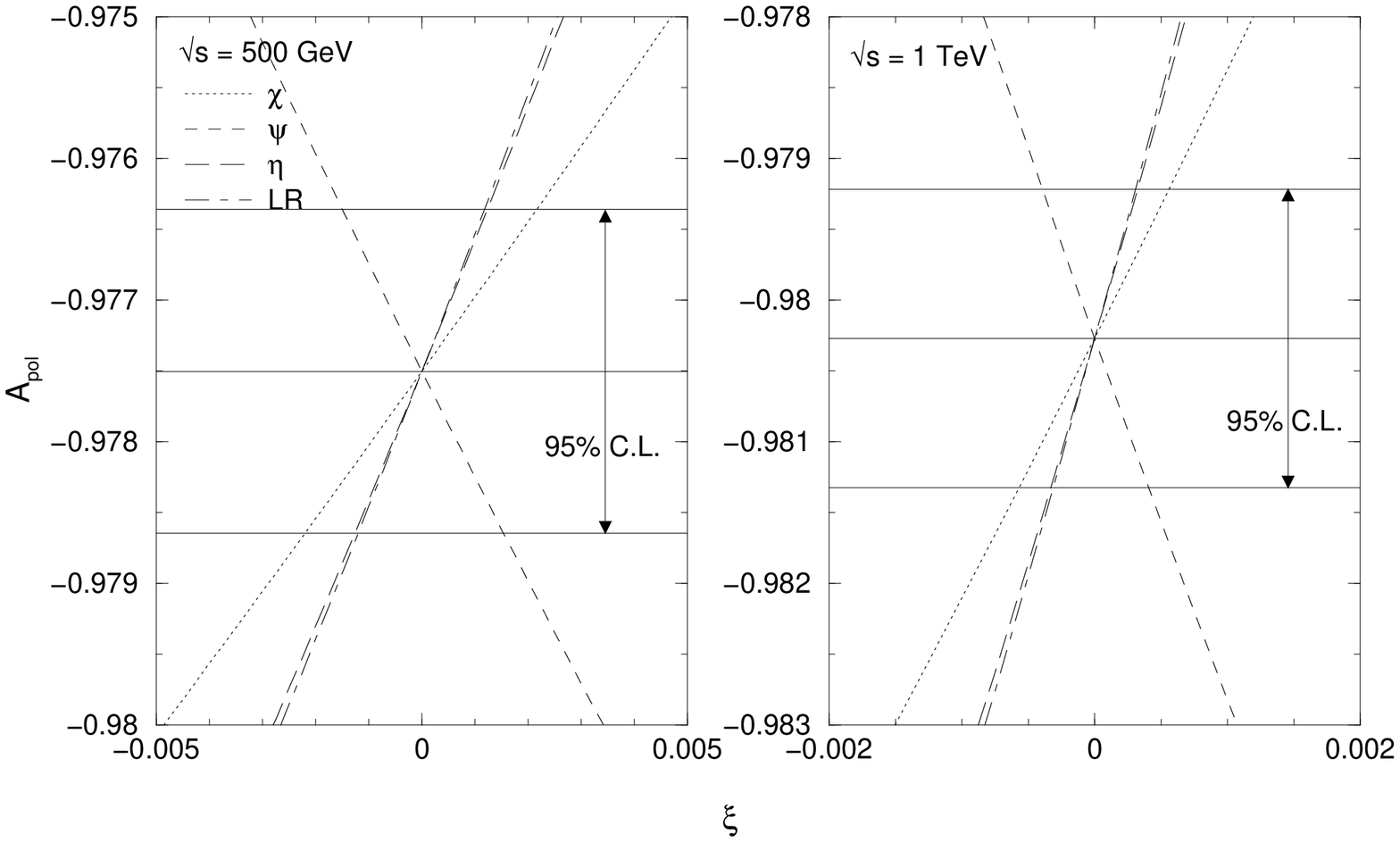,width=19.5cm}}
\end{figure}
\vspace{1.0cm}
\begin{center}
{\large Fig. 7}
\end{center}

\begin{figure}[th]
\centering
\centerline{\epsfig{file=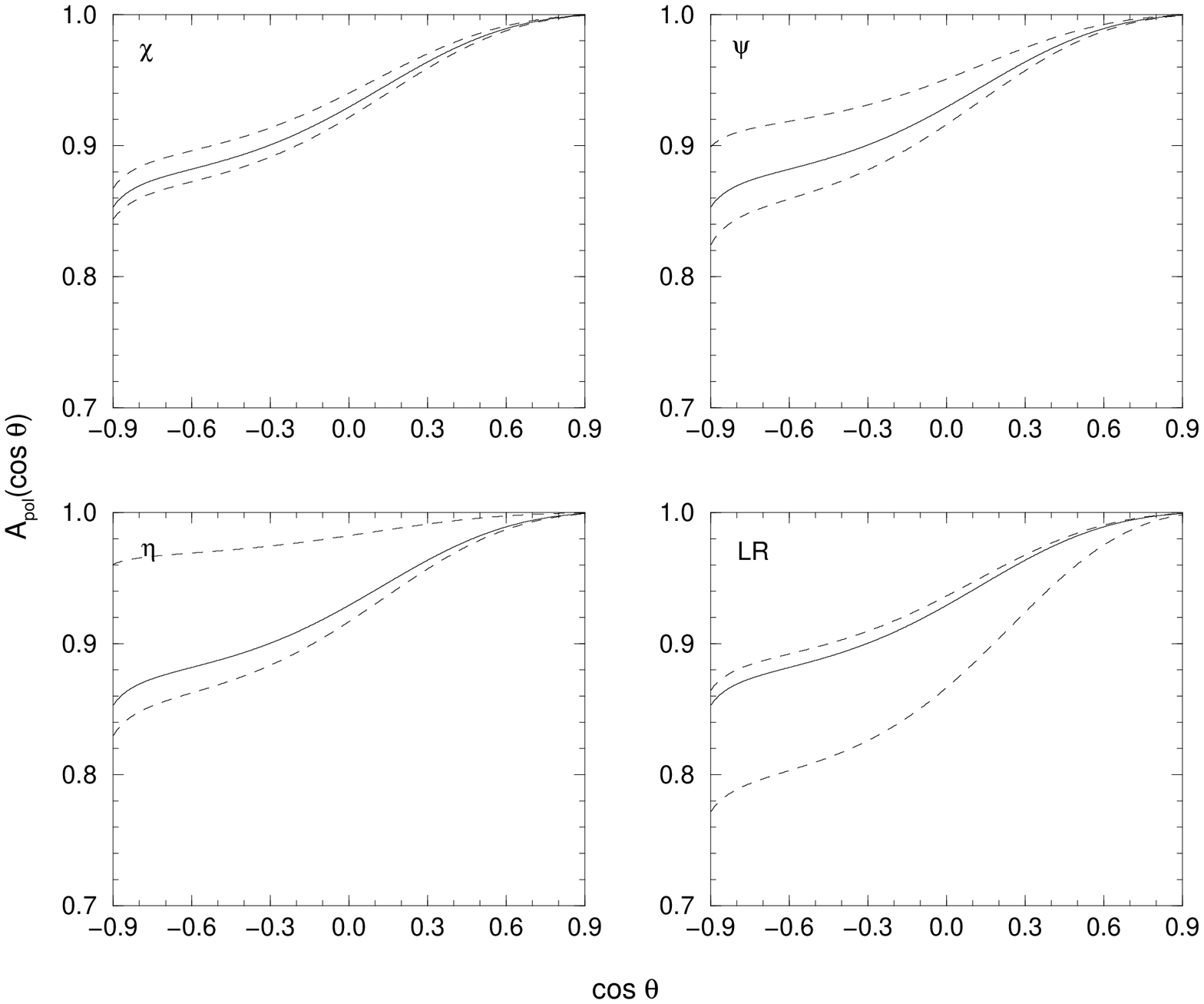}}
\end{figure}
\vspace{1.0cm}
\begin{center}
{\large Fig. 8}
\end{center}

\end{document}